\documentstyle[12pt]{article}
\def\beq{\begin{equation}}   \def\eeq{\end{equation}}
\def\bea{\begin{eqnarray}}   \def\eea{\end{eqnarray}}

\begin{document}

\begin{flushright}
UND-HEP-00-BIG\hspace*{.2em}08\\
\end{flushright}
\vspace{.3cm}
\begin{center} \Large 
{\bf CP Violation and the Cathedral Builders' 
Paradigm}
\footnote{Introductory Lecture given at CPconf2000, Intern. 
Conf. on CP Violation Physics, Sept. 18 - 22, 2000, 
Ferrara, Italy}
\\
\end{center}
\vspace*{.3cm}
\begin{center} {\Large 
I. I. Bigi }\\ 
\vspace{.4cm}
{\normalsize 
{\it Physics Dept.,
Univ. of Notre Dame du
Lac, Notre Dame, IN 46556, U.S.A.} }
\\
\vspace{.3cm}
{\it e-mail address: bigi@undhep.hep.nd.edu } 
\vspace*{0.4cm}

{\Large{\bf Abstract}}
\end{center}

Pointing out the profound and unique nature of 
CP violation, I sketch its basic phenomenology, its  
CKM prescription and QCD technologies relevant for heavy flavour 
physics. After emphasizing the paradigmatic character of the 
establishment of direct CP violation I turn to the future, namely 
indirect searches for New Physics in electric dipole 
moments, $K_{\mu 3}$ decays and charm transitions 
on one hand and in beauty 
decays on the other; these are described as 
exciting adventures with novel challenges 
not encountered before.


\tableofcontents 

\vspace*{1.0cm}

\section{Prologue}

Ferrara is obviously an appropriate site for a conference like ours 
dedicated to cultural and intellectual as well as practical considerations. 
To cite just one historical example as case in point: 
Duke Alfonso I reigned here supporting poets 
and painters while at the same time  
succeeding in making Ferrara's artillery the best in Italy. 
He also married 
Lucrezia Borgia 499 years ago and brought her to 
Ferrara with her celebrated charm and 
beauty; never mind that both were somewhat tainted by scandal. 

In this introductory lecture to the conference I want to 
sketch the big picture on CP violation concerning  
past developments,   
theoretical tools employed and  
future promises and novel challenges. 

I will group the material into three periods:   
\begin{itemize} 
\item  
the `past' -- 
1964 - 1998 -- with the main topics being 
the discovery of CP violation, its theory 
including Penguins and an interlude 
on new theoretical technologies of the '90's, 
namely  lattice QCD and $1/m_Q$ expansions;  
\item   
the `present' -- 1998 - 2002 -- 
with the observation of $\epsilon^{\prime}/\epsilon \neq 0$, 
T violation and CP violation in beauty decays and 
\item   
the `future' -- 2002 - 2015 ff. -- with  
non-mainstream CP violation (electric dipole moments, 
$K_{\mu 3}$ decays,   
$D^0 - \bar D^0$ oscillations and CP violation) as well as   
CKM trigonometry.  
\end{itemize} 
In my summary I will describe the Cathedral Builders' paradigm.

\section{The Past: 1964 - 1998} 

\subsection{CP Violation as a fundamentally new paradigm}

The discovery in 1957 that parity was broken in weak decays 
certainly caused a shock in the community. Yet the latter   
recovered remarkably fast largely due to arguments put 
forward by leading physicists like Landau. They suggested 
one had been hasty in requiring full invariance under parity.  
Invoking somewhat obliquely Mach's principle they 
instead argued in 
favour of CP symmetry 
pairing {\em left}-handed neutrinos 
with {\em right}-handed {\em anti}neutrinos; `left' and `right' 
is then defined in terms `positive' and `negative'. This is 
similar to a German saying that the thumb is `left' on the 
`right' hand: it is as factually correct as it is useless 
since circular. A world of left-handed fermions and right-handed 
antifermions is thus a completely symmetric one. Indeed it was 
found that maximal parity violation in weak 
interactions is matched by maximal 
violation of charge conjugation.

The observation of $K_L \to \pi ^+\pi ^-$ in 1964 was totally 
unexpected by almost all theorists, and they did not give up 
without a fight. Interpretations other than CP violation 
were entertained: the existance of particles $U$ 
escaping detection in $K_L \to \pi ^+\pi ^- [U]$ was postulated
\footnote{It is an argument analogous to Pauli's introduction 
of neutrinos into $\beta$ decay: 
an `invisible' particle is postulated 
to save a conservation law, namely that of energy-momentum 
there and CP here. While this idea worked there, it failed here.} 
; 
cosmological background fields were invoked and even the idea 
of {\em nonlinear effects} in quantum mechanics were floated 
\cite{ROOS}  
-- to no avail! 
The fact that CP invariance appeared to be a `near-miss' -- 
BR$(K_L \to \pi ^+\pi ^-) \sim 0.002 \ll 1$   
in contrast to maximal P violation --   
made it even harder to accept. 
Nevertheless the whole community soon came 
around to accept CP violation as an empirical fact 
\cite{PAIS,CPBOOK}. 

I am telling this story not to poke fun at my predecessors. 
There were very good reasons for theorists' slowness in embracing 
CP violation. For it was clearly realized that CP violation represented 
a more fundamental and radical shift to a new paradigm than 
parity violation. Firstly CP violation means that 
`left'- and `right'-handed can be distinguished in an 
{\em absolute}  
way, independant of any convention concerning the sign of charges. 
This is most obvious from the observation on semileptonic 
$K_L$ decays:
\beq 
\Gamma (K_L \to l^+ \nu _L\pi^-) > 
\Gamma (K_L \to l^- \bar \nu _R\pi^+) \; . 
\eeq  
Secondly based on CPT symmetry CP violation implies T violation, 
i.e. that nature distinguishes between `past' and `future' on the 
{\em microscopic} level. Thirdly one can add (at least in retrospect) 
 that CP violation is a necessary ingredient 
in any effort to understand the baryon number of the Universe 
as a {\em dynamically generated} quantity rather than as a parameter 
reflecting {\em initial conditions}. On a more technical level one 
can point out that CP violation represents the smallest observed 
violation of a symmetry:  
Im$M_{12} \simeq 1.1 \cdot 10^{-8}$ eV or 
Im$M_{12}/m_K \simeq 2.2 \cdot 10^{-17}$. The peculiar role 
of T violation surfaces also through {\em Kramers' Degeneracy} 
\cite{KRAMERS}. 
With the time reversal operator $T$ being {\em anti}unitary, 
$T^2$ has eigenvalues $\pm 1$ meaning the Hilbert space has two 
distinct sectors. It is easily shown that each energy eigenstate 
in the sector with $T^2 = -1$ is at least doubly degenerate. 
This degeneracy is realized in nature through fermionic degrees 
of freedom. I find it quite remarkable that the operator $T$ 
anticipates this option (and the qualitative difference between 
fermions and bosons) through $T^2=-1$ {\em without} any explicit 
reference to spin.

\subsection{Basic CP [\& T] phenomenology and the data in 1998}

Due to CPT symmetry CP and T violation can enter through complex phases 
only. For them to become observable, one needs two different 
amplitudes to contribute coherently. This can be realized 
in different ways: 
\begin{itemize}
\item 
{\em Particle-antiparticle oscillations followed by a decay into 
a common final state:} 

Such asymmetries are often referred to -- with less than Shakespearean 
flourish -- as indirect CP violation. 
The decay rate evolution in proper time then 
differs from a {\em pure exponential}, and the 
difference between CP conjugate transitions becomes a nontrivial 
function of time. Well-known examples are 
$K^0(t) \to \pi ^+\pi^-$ vs. $\bar K^0(t) \to \pi ^+\pi^-$ 
or $B_d(t) \to \psi K_S$ vs. $\bar B_d(t) \to \psi K_S$ with 
\cite{BS} 
\beq  
\Gamma (B_d(t)[\bar B_d(t)] \to \psi K_S) \propto 
e^{-t/\tau (B_d)} \left( 
1 - [+] A {\rm sin}(\Delta m_d t)\right) 
\eeq
Final state interactions (FSI) in general will affect the 
signal, although not in this case. 
On the other hand they are not required and they cannot fake a signal. 
\item 
{\em Direct CP  violation:} 

Within the SM 
they can occur in CKM suppressed modes only.
There are several classes of such effects 
differing in the role played by final state interactions 
(FSI); they all share the feature 
that the signal is independant of the time of decay. 
\begin{itemize}
\item 
{\em Partial width differences:} If the final 
state consists of two pseudoscalar mesons or one 
pseudoscalar and one vector meson, then CP violation can manifest 
itself only in a partial width difference.   
FSI are necessary to transform 
CP violation into an observable. While they cloud the 
numerical interpretation of a signal (or its absence), they cannot 
fake a signal.  
\item 
{\em Final state distributions:} If a final state is more complex, i.e. 
consists of at least three pseudoscalar mesons not forming  
a resonance or of two vector mesons etc., then there are several 
potential layers of dynamical information. There could be 
asymmetries in subregions of a Dalitz plot that are substantially larger 
than when integrated over the whole Dalitz plot. 

Going one step further one can study decays of a particle P into 
four pseudoscalar mesons: $P \to a+b+c+d$. Such a final state allows to 
construct non-trivial {\em T-odd} correlations:
\beq 
C_T\equiv \langle \vec p_a\cdot (\vec p_b \times \vec p_c) \rangle  
\eeq 
with $C_T \to - C_T$ under time reversal. T violation can 
produce $C_T \neq 0$ irrespective of FSI; yet 
$C_T \neq 0$ does not necessarily establish T violation. 
Since T is described by an {\em anti}unitary operator, FSI can 
induce $C_T \neq 0$ with T-invariant dynamics. 
In contrast to the situation with partial widths where 
FSI play the role of a necessary evil, they can act here as 
an imposter.  Yet comparing 
this observable for particle and antiparticle decays and finding 
$C_T + \bar C_T \neq 0$ establishes CP violation.  

The muon polarization transverse to the decay plane in 
$K^+ \to \mu ^+ \pi ^0 \nu$ represents such a T-odd correlation:  
$P_{\perp}(\mu ) = \langle \vec s(\mu ) \cdot 
(\vec p (\mu) \times \vec p (\pi))/
|\vec p (\mu) \times \vec p (\pi)| \rangle$, 
which 
in this case could not be faked realistically 
by final-state interactions and
would  reveal genuine T violation.

\item 
The leading, namely linear term for the energy shift of a system inside a  
weak electric field $\vec E$ is described by a static 
quantity, the electric dipole moment $\vec d$: 
\beq 
\Delta {\cal E} = \vec d \cdot \vec E + {\cal O}(E^2) 
\eeq
For a non-degenerate system with spin $\vec s$ one has 
$\vec d \propto \vec s$; therefore $\vec d \neq 0$ reveals 
T (and P) violation.

\end{itemize}

\end{itemize}

The relevant data read as follows in 1998: 
\begin{itemize}
\item 
\bea 
{\rm BR}(K_L \to \pi ^+ \pi ^-) &\simeq & 2.3 \cdot 10^{-3} \neq 0 \\
\frac{{\rm BR}(K_L \to l^+ \nu \pi ^-)}
{{\rm BR}(K_L \to l^- \nu \pi ^+)} &\simeq & 1.006 \neq 0
\eea
\item 
\beq 
{\rm Re} \frac{\epsilon ^{\prime}}{\epsilon _K} = 
\left\{ 
\begin{array}{l} 
(2.30 \pm 0.65) \cdot 10^{-3} \; \; NA\, 31 \\ 
(0.74 \pm 0.59) \cdot 10^{-3} \; \; 
E\, 731 
\end{array}  
\right. 
\label{DIRECTCP} 
\eeq
\item 
The muon transverse polarization in 
$K^+ \to \mu ^+ \nu \pi ^0$: 
\beq 
{\rm Pol}_{\perp} (\mu )   
 = (-1.85 \pm 3.6) \cdot 10^{-3} 
\eeq 
\item 
Electric dipole moments (EDM) for neutrons and electrons 
\bea 
d_N &<& 9.7 \cdot 10^{-26} \; \; e\, cm \\
d_e &=& (-0.3 \pm 0.8) \cdot 10^{-26} \; \; e \, cm
\eea 
To get an intuitive understanding about the sensitivity 
achieved one can point out that the uncertainty in the 
electron's {\em magnetic} moment is about 
$2 \cdot 10^{-22}$ e cm and thus several orders of magnitude 
larger than the bound on its EDM! 
The bound on the neutron's 
EDM is smaller than its radius by 13 orders of magnitude. 
This corresponds to a relative displacement of an electron and a 
positron spread over the whole earth by less than 1 $\mu$ -- much 
less than the thickness of human hair!

\end{itemize}
The situation in 1998 can then be described as follows: after 34 
years of dedicated experimental work CP violation could still be 
described by a {\em single} number, namely $\epsilon$, the 
situation concerning direct CP violation was in limbo, see 
Eq.(\ref{DIRECTCP}), and no other manifestation had been seen.

\subsection{Theory of CP violation}

Initially it had been suggested that electrodynamics might 
violate CP invariance; yet it was soon cleared of that 
suspicion. There was then no theory of CP violation 
till 1973. The community can be forgiven for not being overly 
concerned about explaining BR$(K_L \to \pi ^+\pi ^-) 
\simeq 0.002$ when there are still infinities arising in the 
theoretical description of weak decays. Yet I find it highly 
remarkable that even after the SM had been formulated as a 
{\em renormalizable} theory by the late 1960's the lack of a theory 
for CP violation was not noticed till 1972 \cite{MOHA}. It is often 
said in response:"Well, we had the superweak model put forward 
by Wolfenstein already in 1964". However I view the superweak 
model 
\cite{SUWE} as a {\em classification} scheme for theories 
rather than a theory 
itself. Whenever one suggests a theory of CP violation, one has to 
analyze whether it provides a dynamical implementation of the 
superweak scenario or not, and to which accuracy it does so. 

In 1973 the celebrated paper by Kobayashi and Maskawa appeared 
in print 
\cite{KM}. It pointed out that the electroweak SM with two full 
families -- i.e. charm included -- conserves CP; secondly it 
demonstrated how different types of New Physics -- more families, 
more Higgs doublets, right-handed currents -- allow CP breaking 
\footnote{It had been noted first by Mohapatra that the SM with two 
families conserves CP. He suggested right-handed currents as 
the origin of CP violation \cite{MOHA}.}
\footnote{One can point out that Kobayashi and Maskawa benefitted from 
some `insider' information: both were working in the Physics 
Department of Nagoya University at that time where, due to 
Sakata and his school, the notion of quarks as 
real rather than merely mathematical objects had been 
readily accepted, as had been the existence of charm due to the 
discovery of Niu \cite{NIU}.}.   
Only one of these variants, namely the one with (at least) three 
families is now referred to as KM ansatz. 

This KM ansatz (in the narrow sense) removes the mystery from the 
apparent `near miss' of CP invariance in $K_L \to \pi \pi$: this 
transition requires the interplay between three families; yet 
the third family is almost decoupled from the first two -- 
not surprisingly (again at least in retrospect) 
considering its much heavier masses. 

A second milestone was reached in the 1970's when the 
relevance of the so-called 
Penguin operators was realized, first in the context of the 
$\Delta I=1/2$ rule 
\cite{ITEP}, then also for allowing for 
$\epsilon ^{\prime}/\epsilon \neq 0$ 
\cite{GILMAN}. Since then the 
treatment of Penguin operators and operator renormalization 
has reached a high level of sophistication 
\cite{BURAS}.

A third milestone is represented by the formulation of the 
`Strong CP Problem'; it still awaits its resolution 
\cite{PECCEI}! 

Another milestone was the realization 
in 1980 that the KM ansatz unequivocally 
predicts large CP asymmetries in some nonleptonic decay 
channels of neutral $B$ mesons like $B_d \to \psi K_S$ 
\cite{CARTER,BS}. 
It was stated explicitely that 
asymmetries could be 1- 20 \% and possible larger -- at a 
time when neither the `long' $B$ lifetime nor the large 
$B_d - \bar B_d$ oscillation rate nor the huge top mass were 
known; at that time a top mass exceeding 60 GeV would have been seen 
as a frivolous notion! 

\subsection{The `unreasonable' success of the CKM description}

The observation of the `long' $B$ lifetime of about 1 psec together 
with the dominance of $b\to c$ over $b\to u$ revealed a hierarchical 
structure in the KM matrix that is expressed in the Wolfenstein 
representation in powers of $\lambda = {\rm tg}\theta _C$. We   
often see plots of the CKM unitarity triangle where the 
constraints coming from various observables appear as    
broad bands. While the latter is often bemoaned, it obscures 
a more fundamental point: the fact that these constraints can be 
represented in such plots at all is quite amazing! 
The quark box 
{\em without} GIM subtraction yields a value for 
$\Delta m_K$ exceeding the experimental 
number by more than a factor of thousand; it is the GIM mechanism 
that brings it down to within a factor of two or so of experiment. 
The GIM subtracted quark box for $\Delta M_B$ coincides 
with the data again within a factor of two. Yet if the 
beauty lifetime were of order $10^{-14}$ sec while 
$m_t \sim 180$ GeV it would exceed it by 
an order of magnitude; on the other hand it would undershoot by an order 
of magnitude if $m_t \sim 40$ GeV were used with 
$\tau (B) \sim 10^{-12}$ sec; i.e., the observed value can be 
accommodated because a tiny value of $|V(td)V(ts)|$ is offset 
by a large $m_t$. 

This amazing success is repeated with $\epsilon$. Over the last 
25 years it could always be accommodated (apart from 
some very short periods of grumbling mostly off the record) 
whether the {\em correct} set [$m_t = 180$ GeV with $|V(td)|\sim 
\lambda ^3$, 
$|V(ts)|\sim \lambda ^2$] or the {\em wrong} one 
[$m_t = 40$ GeV with $|V(td)|\sim \lambda ^2$, 
$|V(ts)|=\lambda$] were used. Yet both 
$m_t = 180$ GeV with $|V(td)|=\lambda ^2$, 
$|V(ts)|=\lambda$ as well as 
$m_t = 40$ GeV with $|V(td)|=\lambda ^3$, 
$|V(ts)|=\lambda ^2$ would have lead to a clear inconsistency! 

Thus the phenomenological success of the CKM description has to be 
seen as highly nontrivial or `unreasonable'. This cannot have 
come about by accident -- there must be a profound reason.

\subsection{Experimental developments}

As we all know (and will be reminded of during this conference) 
there is a worldwide and dedicated program of $B$ physics underway 
now. It has benefitted tremendously from experimental developments 
that could hardly be anticipated in 1980. 

Driven by the demads of charm physics the technology for microvertex 
detectors was developed that resolves secondary decay 
vertices of charm and beauty states giving meaning 
to the term `long' $B$ lifetime; it also allows to track  
$B_d - \bar B_d$ oscillations which were first discovered in 1987  
at a rate close to the decay rate. Various methods for `opposite-side' and 
`same-side' flavour tagging were pioneered in charm studies. Finally 
the concept of asymmetric $e^+e^-$ colliders and detectors for them  
(something that at first 
had to be seen as quite frivolous to put it mildly) was born -- 
and realized!

\subsection{New QCD technologies of the 1990's}

Since we have to study the decays of quarks bound inside hadrons, 
we have to deal with nonperturbative dynamics 
\footnote{Since top quarks decay before they can hadronize, 
their interactions can be treated perturbatively 
\cite{DOK}.} 
-- 
a problem that in general has not been brought under theoretical 
control. Yet we can employ various theoretical technologies 
that allow to treat nonperturbative effects in special situation: 
\begin{itemize}
\item 
For {\em strange} hadrons where $m_s \leq \Lambda _{QCD}$ 
one invokes chiral perturbation theory. 
\item 
For {\em beauty} hadrons with $m_b \gg \Lambda _{QCD}$ one can 
employ $1/m_b$ expansions in various incarnations; they should provide 
us with rather reliable results, whenever an operator product expansion 
can be applied \cite{HQT}. 
\item 
It is natural to extrapolate such expansions down to the charm 
scale; this has to be done with considerable caution, though:  
while the charm quark mass does exceed ordinary hadronic mass 
scales, it does not do so by a large amount.  
\item 
Lattice QCD on the other hand is most readily set up at ordinary 
hadronic scales; from those one extrapolates {\em down} towards the chiral 
limit (which represents a nontrivial challenge \cite{SONI}) and 
{\em up} to the charm scale and beyond.

\end{itemize}

\noindent Let me add a few more specific comments: 
 
Lattice QCD, which originally had been introduced to prove confinement 
and bring hadronic spectroscopy under computational control is now making 
major contributions to heavy flavour physics. 
This can be illustrated with very recent results on decay constants 
where the first {\em un}quenched results (with two dynamical 
quark flavours) have become available \cite{KENWAY}. 
\begin{itemize}
\item 
\beq 
f(D_s) = 
\left\{ 
\begin{array}{l} 
240 \pm 4 \pm 24, \, 275 \pm 20 \;
{\rm MeV}, \; {\rm lattice\, QCD}\\ 
269 \pm 22 \; {\rm MeV}, \; {\rm world\, average\, of\, data}  
\end{array}  
\right. 
\eeq
\item 
\bea 
f(B) &=& 190 \pm 6 \pm 20 ^{+9}_{-0} \; {\rm MeV}, 
\; {\rm lattice\, QCD}\\
f(B_s) &=& 218 \pm 5 \pm 26 ^{+9}_{-0} \; {\rm MeV}, 
\; {\rm lattice\, QCD}
\eea
\end{itemize}

The $1/m_Q$ expansions have become more refined and reliable 
qualitatively as well as quantitatively: 
\begin{itemize}
\item 
The $b$ quark mass has been extracted 
from data by different groups; 
their findings, when expressed in terms of the 
socalled `kinetic' mass, read as follows: 
\beq 
m_b^{\rm kin} (1\, {\rm GeV}) = 
\left\{ 
\begin{array}{l} 
4.56 \pm 0.06  \; \; 
{\rm GeV} \; \;  \cite{MEL}, \\  
4.57 \pm 0.04  \; \; 
{\rm GeV} \; \;  \cite{HOANG}, \\  
4.59 \pm 0.06  \; \; 
{\rm GeV} \; \;  \cite{SIGNER}
\end{array}  
\right.
\eeq
The error estimates of 1 - 1.5 \% might be overly optimistic (as it 
often happens), but not foolish. Since all 
three analyses use basically the same input from the 
$\Upsilon (4S)$ region, they could suffer from a common 
systematic uncertainty, though. 
\item 
For the form factor 
describing $B\to l \nu D^*$ at zero recoil
 one has the following results:
\beq 
F_{D^*}(0) = 
\left\{ 
\begin{array}{l} 
0.89 \pm 0.08  \; \;       \cite{URI1}, \\ 
0.913 \pm 0.042 \; \; \cite{BABARBOOK}, \\  
0.935 \pm 0.03 \; \; \cite{LAT} 
\end{array}  
\right. 
\eeq 
where the last number has been obtained in lattice QCD. 
\end{itemize}

There is a natural feedback between lattice QCD and $1/m_Q$ 
expansions: by now both represent mature technologies that 
are defined in Euclidean rather than Minkowskian space; 
they share some expansion parameters, while differing in others; 
lattice QCD can evaluate hadronic matrix elements that serve 
as input parameters to $1/m_Q$ expansions. 

It has been accepted for a long time now that heavy flavour decays 
can serve as high {\em sensitivity} probes for New Physics. I feel 
increasingly optimistic that our tools are and will be such that 
that they will provide us even with high {\em accuracy} probes!

\subsection{Expectations and predictions 1998}

The observed hierarchy in the CKM parameters 
\beq 
|V(ub)|^2 \ll |V(cb)|^2 \ll |V(cd)|^2 
\eeq
tells us that the CKM matrix can conveniently be described by 
the Wolfenstein parametrization in powers of 
$\lambda = {\rm tg}(\theta _C)$: 
\beq 
V_{CKM} = \left( 
\begin{array}{ccc} 
V(ud) & V(us) & V(ub) \\ 
V(cd) &V(cs) & V(cb) \\ 
V(td) & V(ts) & V(tb) 
\end {array} 
\right) = 
\left( 
\begin{array}{ccc} 
1 & {\cal O}(\lambda ) & {\cal O}(\lambda ^3) \\ 
{\cal O}(\lambda ) & 1 & {\cal O}(\lambda ^2) \\ 
{\cal O}(\lambda ^3) & {\cal O}(\lambda ^2) & 1 
\end {array} 
\right) 
\label{WOLF} 
\eeq
More specifically PDG2000 states as 90 \% C.L. ranges   
\beq 
|V_{CKM}| = 
\left( 
\begin{array}{ccc} 
0.9750 \pm 0.0008 & 0.223 \pm 0.004 & 0.003 \pm 0.002 \\ 
0.222 \pm 0.003 & 0.9742 \pm 0.0008 & 0.040 \pm 0.003 \\ 
0.009 \pm 0.005 & 0.039 \pm 0.004 & 0.9992 \pm 0.0002 
\end {array} 
\right)
\label{CKM3}
\eeq
Without imposing three-family unitarity that is implicit in the 
Wolfenstein representation PDG2000 lists numbers that in particular 
for the top couplings are much less restrictive:  
\beq 
|V_{CKM}| = 
\left( 
\begin{array}{cccc} 
0.9735 \pm 0.0013 & 0.220 \pm 0.004 & 0.003 \pm 0.002 & ...\\ 
0.226 \pm 0.007 & 0.880 \pm 0.096 & 0.040 \pm 0.003 & ...\\ 
0.05 \pm 0.04 & 0.28 \pm 0.27 & 0.5 \pm 0.49 & ... \\
... & ... & ... & ... 
\end {array} 
\right)
\label{CKM4}
\eeq
I would like to add two comments here: 
(i) The brandnew CLEO number for $|V(cb)|$ from 
$B\to l \nu D^*$ -- 
$|V(cb)F_{D^*}(0)| = (42.4 \pm 1.8 \pm 1.9)\times 10^{-3}$ 
\cite{CINABRO} -- 
falls outside the 90\% C.L. range stated by PDG2000 
for the expected values of 
$F_{D^*}(0)$.  (ii) The OPAL collaboration has presented 
a new {\em direct} determination of $|V(cs)|$ from 
$W\to H_c X$: $|V(cs)| = 0.969 \pm 0.058$ 
\cite{OPAL}. 

With these input values one can make predictions on CP asymmetries, 
at least in principle and to some degree. I will confine myself 
to a few more qualitative comments since these issues will be discussed 
in great detail in subsequent talks at this conference. 
\begin{itemize}
\item 
If there is a single CP violating phase 
$\delta$ as is the case in the KM 
ansatz one can conclude based on the $\Delta I = 1/2$ rule: 
$\epsilon ^{\prime}/\epsilon \leq 1/20$. The large top mass 
-- $m_t  \gg M_W$ -- enhances the SM prediction for $\epsilon$ 
considerably more than for $\epsilon ^{\prime}$ for a 
given $\delta$ and therefore on quite general grounds 
\beq 
\epsilon ^{\prime}/\epsilon \ll 1/20
\eeq
\item 
Of course the KM predictions made employed much more 
sophisticated theoretical reasoning. Before 1999 they tended to 
yield -- with few exceptions 
\cite{FABB} -- values not exceeding $10^{-3}$ 
due to sizeable cancellations between different contributions. 
\item 
Once the CKM matrix exhibits the {\em qualitative} 
pattern given in Eq.(\ref{WOLF}), 
it neccessarily follows that certain $B_d$ decay channels will 
exhibit CP asymmetries of order unity. To be more specific one can 
combine what is known about $V(cb)$, $V(ub)$, $V(ts)$ and $V(td)$ 
from semileptonic $B$ decays, $B_d - \bar B_d$ oscillations and 
bounds on $B_s - \bar B_s$ oscillations with or without using 
$\epsilon$ to construct the CKM unitarity triangle 
describing $B$ decays. A crucial question to which I will return 
later centers on the proper treatment of theoretical uncertainties. 
A typical example is \cite{PARODI}: 
\bea 
{\rm sin} 2\phi _1[\beta] &=& 0.716 \pm 0.070 \\ 
{\rm sin} 2\phi _2[\alpha] &=& - 0.26 \pm 0.28 
\eea 
\end{itemize}

\section{The Present: 1999 - $\sim$ 2002}

A new period began in 1999 when direct CP violation became established 
in $K_L$ decays and the new $B$ factories started up. I expect those 
$B$ factories to have established CP violation in at 
least one $B$ decay mode by 2002. 

I will confine myself to a few brief comments 
on this present period since that is 
the subject of this conference, and I do not want to overengage in 
poaching. 
 
\begin{itemize}
\item 
There can no longer be any doubt that direct CP violation has been 
observed in $K_L$ decays although its actual strength is not precisely 
known yet. It is a discovery of the first rank irrespective of what 
theory says or does not say. 

As I had argued before a rather small, but 
nonzero value is a natural expectation of the KM ansatz. 
To go beyond such a qualitative statement, one has to evaluate 
hadronic matrix elements; apparently one had 
underestimated the complexities in this task. 
One intriguing aspect in this development 
is the saga of the $\Delta I=1/2$ rule: formulated in a compact way 
\cite{DELTARULE} it 
was originally expected to find a simple dynamical explanation; several 
enhancement factors were indeed found, but the observed enhancement 
could not be reproduced in a convincing manner; this problem was then 
bracketed for some future reconsideration and it was argued that 
$\epsilon ^{\prime}/\epsilon$ could be predicted while ignoring the 
$\Delta I=1/2$ rule. Some heretics -- `early' ones  
\cite{SANDA} and `just-in-time' ones \cite{TRIESTE} -- 
however argued that only approaches that reproduce the observed 
$\Delta I=1/2$ enhancement can be trusted to yield a realistic 
estimate of $\epsilon ^{\prime}/\epsilon$. 

\item 
It is often 
alleged that 
CPT invariance can boast of impressive experimental verification 
as expressed 
through the bound 
$|M(K^0) - M(\bar K^0)|/M(K) = (0.08 \pm 5.3) \cdot 10^{-19}$. 
However one might as well have divided this difference by the mass of an 
elephant since {\em intrinsically} the kaon mass is hardly more 
related to the $K - \bar K$ mass splitting than the elephant's mass. 

To put it differently: since this CPT breaking is expressed through 
a mass difference, one needs another 
dimensional quantity as yardstick.  This can be provided by 
Im$M_{12}$ expressing CP violation in the mass matrix: 
\beq 
|M(K^0) - M(\bar K^0)| < 2.5 \cdot 10^{-10} \; {\rm eV} 
\; \; \Leftrightarrow \; \; 
{\rm Im}M_{12} \simeq 10^{-8} \; {\rm eV} \; ; 
\eeq
i.e., CPT breaking still could be as `large'as a few percent of the 
observed CP violation! 

Our belief in 
CPT invariance is of course based much more on `dogma', i.e. theory, 
than empirical facts. For it 
is an almost inescapable consequence of {\em local} 
quantum field theories based on canonized assumptions like 
Lorentz invariance, the existence of a unique vacuum state and 
weak local commutativity obeying the `right' statistics. 
Some explicit examples of CPT breaking theories have been 
given, but they are highly contrived and unattractive 
\cite{TOD,TUMB}. 

The new interest in experimental studies of CPT symmetry is 
fed by two more recent developments 
\cite{KOST}: 
\begin{itemize}
\item 
Novel tests of CPT as well as {\em linear} quantum mechanics 
can be performed at the $\Phi$ and beauty factories DA$\Phi$NE, 
BABAR and BELLE respectively by harnessing EPR correlations 
\cite{EPR}. 
\item 
Superstring theories are intrinsically {\em non}local thus vitiating 
one of the central axioms of the CPT theorem. Furthermore gravity 
could induce CPT breaking either as a true symmetry violation or as 
a background effect due to the preponderance of matter over antimatter 
in our corner of the universe. It would then be not unreasonable to 
expect CPT asymmteries to scale like a positive power of 
$E/M_{Planck}$; if that power were unity one would guestimate 
$|M(K^0) - M(\bar K^0)| \sim M(K)/M_{Planck} \sim 10^{-19}$!

\end{itemize}
 
\item 
Although CP violation implies T violation due to the CPT theorem, 
I consider it
highly significant that more  direct evidence has been obtained through the 
`Kabir test': CPLEAR has found \cite{CPLEAR} 
\beq 
A_T \equiv 
\frac{\Gamma (K^0 \to \bar K^0) - \Gamma (\bar K^0 \to K^0)}
{\Gamma (K^0 \to \bar K^0) + \Gamma (\bar K^0 \to K^0)} = 
(6.6 \pm 1.3 \pm 1.0)\cdot 10^{-3} 
\eeq
versus the value $(6.54 \pm 0.24)\cdot 10^{-3}$ inferred from 
$K_L \to \pi^+\pi^-$. Of course, some assumptions still 
have to be made, namely that {\em semileptonic} $K$ decays obey 
CPT or that the Bell-Steinberger relation is satisfied with 
{\em known} decay channels only. Avoiding both assumptions 
one can write down an 
admittedly contrived scheme where the CPLEAR data are  
reproduced {\em without} T violation; the price one pays is a large CPT 
asymmetry $\sim {\cal O}(10^{-3})$ in 
$K^{\pm} \to \pi ^{\pm}\pi ^0$ \cite{TBS}. 

\item 
KTeV and NA48 have analyzed the rare decay 
$K_L \to \pi^+\pi^- e^+e^-$ and found a large {\em T-odd} 
correlation between the $\pi^+\pi^-$ and $e^+e^-$ planes in 
full agreement with predictions \cite{SEHGAL}. 
Let me add just two comments here: (i) This agreement cannot be 
seen as a success for the KM ansatz. Any scheme reproducing  
$\eta _{+-}$ will do the same. (ii) The argument that strong final state 
interactions (which are needed to generate a T odd correlation 
above 1\% with T invariant dynamics) cannot affect the 
relative orientation 
of the $e^+e^-$ and $\pi ^+\pi ^-$ planes fails on the 
quantum level \cite{TBS}. 
 
The effect found represents a true CP asymmetry. Yet if one is 
sufficiently determined, it still could be attributed to CP and 
CPT breaking that leaves T invariant. A more detailed discussion 
of these subtle points is given in Sehgal's talk.

\end{itemize}

\section{The Future: 2005 - 2015 ff.}

Considerable circumstantial evidence has been accumulated 
that the SM is incomplete. There are (at least) four central mysteries 
at the basis of flavour dynamics: 
\begin{itemize}
\item 
Why is there a family 
structure relating quarks and leptons? 
\item 
Why is there more than 
one family, why three, is three a fundamental parameter? 
\item 
What is the origin of the observed pattern in the quark masses 
and the  
CKM parameters? This pattern can hardly have come about by accident. 
\item 
Why are neutrinos massless -- or aren't they? 
\end{itemize}
To a large degree studying flavour dynamics represents an indirect 
or high sensitivity search for New Physics. 

The significance of the dates is that by 2005 or so the data flow 
on $B$ decays will increase very significantly allowing for 
precision measurements and by about 2015 such precision measurements 
from the $e^+e^-$ factories and from hadronic colliders will have been 
made. 

\subsection{Searching for qualitative discrepancies}

$\Delta S=1,2$ dynamics have provided several examples of revealing the 
intervention of features that represented New Physics 
{\em at that time}; 
it thus has been instrumental in the evolution of the SM. This happened 
through the observation of `qualitative' discrepencies; i.e., 
rates that were expected to vanish did not, or rates were 
found to be smaller than expected by several orders of magnitude. 
Such an indirect search for New Physics can be characterised as a 
`King Kong' scenario: "One might be unlikely to encounter King 
Kong; yet once it happens there can be no doubt that one has 
come across someting out of the ordinary". Such a situation 
can be realized for charm and $K_{\mu 3}$ decays and EDMs. 

\subsubsection{$P_{\perp}(\mu )$ in $K^+ \to \mu ^+ \pi ^0 \nu$}

With $P_{\perp}(\mu ) \sim 10^{-6}$ in the SM, it would also 
reveal New Physics that has to involve chirality breaking weak 
couplings: $P_{\perp}(\mu ) \propto {\rm Im}\xi$, where 
$\xi \equiv f_-/f_+$ with 
$f_-[f_+]$ denoting the chirality violating [conserving] 
decay amplitude. There is an on-going experiment at KEK 
(KEK-E 246) aiming at 
a sensitivity for $P_{\perp}(\mu )$ of $10^{-3}$ or better.

\subsubsection{EDM's}

With the KM scheme predicting unobservably tiny effects 
(with the only exception being the `strong CP' problem) --  
namely $d_{N,e} < 10^{-30}$ e cm --  
and many New Physics scenarios yielding 
$d_{neutron}$, $d_{electron}$ $\geq 10^{-27}$ ecm, this is truly a 
promising zero background search for New Physics! The next round of 
experiments is aiming at $10^{-28}$ e cm for $d_N$ and 
$10^{-30}$ e cm for $d_e$. 

The game one is hunting is actually much more numerous, since many 
effects from the domain of nuclear physics can be employed here. 
These intriguing possibilities are discussed in Hinds' talk.

\subsubsection{$D^0$ Oscillations \& CP Violation}

It is often stated that $D^0$ oscillations are slow and 
CP asymmetries tiny within the SM and that therefore their analysis 
provides us with zero-background searches for New Physics. 

Oscillations are described by the normalized mass and width 
differences:  
$x_D \equiv \frac{\Delta M_D}{\Gamma _D}$,   
$y_D \equiv \frac{\Delta \Gamma}{2\Gamma _D}$.  
A conservative SM estimate yields $x_D$, $y_D$ $\sim 
{\cal O}(0.01)$. Stronger bounds have appeared in the literature, 
namely that the contributions from the 
operator product expansion (OPE) are completely insignificant 
and that long distance contributions {\em beyond} the OPE provide the 
dominant effects yielding $x_D^{SM}$, $y_D^{SM}$ 
$\sim {\cal O}(10^{-4} - 10^{-3})$. A recent detailed analysis 
\cite{BUOSC} revealed 
that a proper OPE treatment reproduces also such long distance 
contributions with  
\beq 
x_D^{SM}|_{OPE}, \, y_D^{SM}|_{OPE} \sim {\cal O}(10^{-3}) 
\eeq   
and that $\Delta \Gamma $, which is generated from  
on-shell contributions, is -- in contrast to $\Delta m_D$
-- insensitive to New Physics while on the other hand more susceptible 
to violations of (quark-hadron) duality. 

Four experiments have reported new data on $y_D$: 
\bea
 y_D = (0.8 \pm 2.9 \pm 1.0) \% \; \; {\rm E791}&,& \; 
(3.42 \pm 1.39 \pm 0.74) \% \; \; {\rm FOCUS} \\
y_D = (1.0^{+3.8 + 1.1}_{-3.5-2.1})\% \; \; {\rm BELLE} &,& \; 
y_D^{\prime} = (-2.5 ^{+1.4}_{-1.6} \pm 0.3)\% \; \; {\rm CLEO}
\eea
E 791 and FOCUS compare the lifetimes for two different channels, 
whereas CLEO fits a general lifetime evolution to 
$D^0(t) \to K^+\pi ^-$; its $y_D^{\prime}$ depends on the strong 
rescattering phase between $D^0 \to K^-\pi^+$ and 
$D^0 \to K^+\pi^-$ and therefore could differ substantially from 
$y_D$ -- even in sign \cite{PETROV} -- if that phase were 
sufficiently large. 
The FOCUS data contain a suggestion that the lifetime 
difference in the 
$D^0 - \bar D^0$ complex might be as large as ${\cal O}(1\% )$. 
{\em If} $y_D$ indeed were $\sim 0.01$, two scenarios could arise 
for the mass difference. If $x_D \leq {\rm few} \times 10^{-3}$ 
were found, one would infer that the $1/m_c$ expansion yields a 
correct semiquantitative result while blaming the large value for 
$y_D$ on a sizeable and not totally surprising violation of 
duality. If on the other hand $x_D \sim 0.01$ would emerge, we would face 
a theoretical conundrum: an interpretation ascribing this to New 
Physics would hardly be convincing since $x_D \sim y_D$. A more sober 
interpretation would be to blame it on duality violation or on the 
$1/m_c$ expansion being numerically unreliable. Observing 
$D^0$ oscillations then would not constitute a `King Kong' 
scenario. 

Searching for {\em direct} CP violation in 
Cabibbo suppressed $D$ decays as a sign for New Physics would also 
represent a very complex challenge: within the KM description one expects 
to find some asymmetries of order 0.1 \%; yet it would be hard 
to conclusively rule out some more or less accidental enhancement due to a 
resonance etc. raising an asymmetry to the 1\% level. 

The only clean environment is provided by CP violation involving 
$D^0$ oscillations, like in $D^0(t) \to K^+ K^-$ and/or 
$D^0(t) \to K^+ \pi ^-$. For the asymmetry would depend 
on the product sin$(\Delta m_D t) \cdot {\rm Im}
[T(\bar D\to f)/T(D\to \bar f)]$: with both factors being 
$\sim
{\cal O}(10^{-3})$ in the SM one predicts a practically zero 
effect. Yet New Physics scenarios can induce signals as large 
as order 1 percent for $D^0(t) \to K^+ K^-$ and even larger for 
$D^0(t) \to K^+ \pi^-$

\subsection{Beauty physics}

There are several different layers of beauty transitions driven by 
$\Delta B = 1 \& 2$ dynamics, and they are realized in a multitude 
of different channels. Thus there are many opportunities for finding 
New Physics. This can be expressed also by pointing out that there 
are actually six KM unitarity triangles with several of their angles 
affecting CP asymmetries in $B$ decays 
\cite{TRIANGLES}. One is particularly intriguing, 
namely the angle that controls the asymmetry in 
$B_s(t) \to \psi \phi$ or $B_s(t) \to \psi \eta$: it is 
${\cal O}(\lambda ^2)$ 
\cite{BS} and about 2 \%. Yet New Physics could very 
possibly raise it even by an order of magnitude. These modes 
could thus reveal what I have referred to as a {\em qualitative} 
discrepancy. 

Yet the more typical situation is that the expected asymmetry is 
already large. Thus one is faced with a novel challenge: can one be 
confident of having established the presence of New Physics when 
the difference between the expected and the observed signal is much 
less than an order of magnitude? To be more specific: assume 
one predicts an asymmetry of 40 \%, yet observes - 40 \%, can one 
be certain of New Physics? What about if one observes 60 \%? 50\%? 
Interpreting such {\em quantitative} 
discrepancies represents a completely new challenge 
which we have not faced before. 

\subsubsection{Quantitative discrepancies}

I expect a number of asymmetries to be measured within a few 
percent uncertainty, although this is easier said than done. 
The crucial question is whether this experimental sensitivity 
can be exploited theoretically. I am optimistic that the 
value of $|V(cb)|$ will be known to better than 5 \% over 
the next few years, $|V(ub)|$ and maybe also $|V(td)|$ 
to within 5 - 10 \% in the long run. This would imply that 
one could make KM predictions for CP 
asymmetries with typically 5 \% accuracy. However -- 
what exactly does one mean by theoretical uncertainty? 

In my judgement there is no unambigous answer in general. For I view 
theoretical uncertainties to be mostly in the class of 
systematic errors, which are notoriously hard to evaluate. 
Furthermore no uniform standard has been established among 
theorists for stating a range for a theoretical uncertainty. 
My understanding behind
quoting the latter is the  following: "I would be very surprised if the 
true value would fall outside the stated range." Such a 
statement is obviously hard to quantify. 

An extensive literature on how to evaluate them has emerged over the last 
two years in particular 
(see, for example, \cite{PARODI}). 
It seems to me that the passion of
the debate  has overshadowed the fact that a lot of learning has happened. For 
example it is increasingly understood that any value within a stated 
range has to be viewed as equally likely. While concerns are 
legitimate that some actors might be overly 
aggressive in stating constraints on the KM triangle, it would be 
unfair to characterize them as silly. I also view it 
as counterproductive to bless one approach while anathematizing all others 
`ex cathedra'. I believe many different paths should be pursued 
since "good decisions come from experience that often is learnt from  
bad decisions". 

There is one feature of the `scanning method' \cite{BABARBOOK} which 
I find particularly informative since it enhances the 
transparency of the underlying information. For each of the 
theoretical input quantities which reflect the size of hadronic 
matrix elements one picks one acceptable value; with this set 
one deduces constraints on the unitarity triangle from the 
available data and circles the resulting allowed area by a line. 
Then one selects another set of acceptable input values and 
repeats the procedure etc. Such a display reflects the overall 
uncertainty through the area covered by the union of these subdomains; 
at the same time it separates the impact of the theoretical and 
experimental uncertainties and that is a major help in understanding 
the origins of the constraints.

Our most powerful weapon for controlling theoretical uncertainties 
will again be  
{\em over}determining basic quantities by extracting their values from more 
than one independant measurement. In this respect the situation is 
actually more favourable in $B$ than in $K$ decays since there 
are fewer free parameters {\em relative} to the number of available 
decay modes. Once the investment has been made to collect the huge 
number of decays required 
to obtain a sufficient number of the transitions  
of primary interest -- say $B_d \to \psi K_S \to (l^+l^-)_{\psi} 
(\pi ^+\pi ^-)_{K_S}$ -- then we have also a slew of many other 
channels that can act as cross checks or provide us with information 
about hadronization effects etc.  Finally one should clearly distinguish 
the goal one has in mind: does one want to state the most likely 
expectation -- or does one want to infer the presence of New Physics 
from a
discrepancy between  expectations and data? The latter goal is of 
course much more ambitious where for once being conservative is a virtue!

\section{The Cathedral Builders' Paradigm}
\subsection{The Paradigm}

The dynamical ingredients for numerous and multi-layered 
manifestations of CP and T violations do exist or are likely to exist. 
Accordingly one searches 
for them in many phenomena, namely in  
\begin{itemize}
\item 
the neutron electric dipole moment probed with ultracold 
neutrons at ILL in Grenoble, France; 
\item 
the electric dipole moment of electrons studied through the 
dipole moment of atoms at Seattle, Berkeley and Amherst in the US; 
\item 
the transverse polarization of muons in 
$K^- \to \mu ^- \bar \nu \pi ^0$ at KEK in Japan; 
\item 
$\epsilon ^{\prime}/\epsilon _K$ as obtained from $K_L$ 
decays at FNAL and CERN and soon at DA$\Phi$NE in Italy; 
\item 
in decay distributions of hyperons at FNAL; 
\item 
likewise for $\tau$ leptons at CERN, the beauty factories and BES 
in Beijing; 
\item 
CP violation in the decays of charm hadrons produced 
at FNAL and the beauty factories; 
\item 
CP asymmetries in beauty decays at DESY, at the beauty 
factories at Cornell, SLAC and KEK, at the FNAL collider and 
ultimately at the LHC. 

\end{itemize} 
A quick glance at this list already makes it clear 
that frontline research on this topic 
is pursued at all high energy labs in the world -- and then some; 
techniques from several different branches of physics -- 
atomic, nuclear and high energy physics -- are harnessed in 
this endeavour together with a wide range of set-ups. 
Lastly, experiments are performed at the lowest temperatures 
that can be realized on earth -- ultracold neutrons -- and at the 
highest -- in collisions produced at the LHC. And all of that dedicated 
to one profound goal. 
At this point I can explain what I mean by the term 
"Cathedral Builders' Paradigm". 
The building of cathedrals required interregional collaborations, 
front line technology (for the period) from many different fields 
and commitment; it had to be based on solid foundations -- and 
it took time. The analogy to the ways and needs of high energy 
physics are obvious -- but it goes deeper than that. 
At first sight a cathedral looks 
like a very complicated and confusing structure with something 
here and something there. Yet further scrutiny reveals that 
a cathedral is more appropriately 
characterized as a complex rather than a complicated 
structure, one that is multi-faceted and multi-layered -- 
with a coherent theme! One cannot (at least for 
first rate cathedrals) remove any of its elements 
without diluting (or even destroying) its technical soundness and 
intellectual message. Neither can one in our efforts to come to grips 
with CP violation!  

\subsection{Outlook}
I want to start with a statement about the past: 
{\em The comprehensive study of kaon and hyperon physics 
has been instrumental in guiding us to the Standard Model.}  
\begin{itemize}
\item 
The $\tau -\theta $ puzzle led to the realization that parity is not 
conserved in nature. 
\item 
The observation that the production rate exceeded the decay rate 
by many orders of magnitude -- this was the origin of the 
name `strange particles' -- was explained through postulating 
a new quantum number -- `strangeness' -- conserved by the strong, 
though not the weak forces. This was the beginning of the second 
quark family. 
\item 
The absence of flavour-changing neutral currents was incorporated 
through the introduction of the quantum number `charm', which 
completed the second quark family. 
\item 
CP violation finally led to postulating yet another, the third 
family. 
\end{itemize}
All of these elements which are now essential pillars of the Standard 
Model were New Physics at {\em that} time! 

I take this historical 
precedent as clue that a detailed, comprehensive and thus 
neccessarily long-term program on the dynamics of heavy flavours 
-- on the quark as well as lepton side -- in general and on CP 
violation in particular will lead to a 
new paradigm, a {\em new} Standard Model. For we are addressing 
the problem of fermion mass generation -- a 
central mystery in our present SM. Such studies are of fundamental 
importance, they will teach us lessons that cannot be obtained 
any other way and cannot become obsolete. 

It will not be an easy journey, nor will it be short, but we are at 
the beginning of an exciting adventure  -- and we are highly privileged 
to participate!

\vskip 3mm  
{\bf Acknowledgements} 

Ferrara with its austere beauty and culinary charm has provided us 
with the perfect setting for an inspiring meeting. Prof. Savrie and his 
colleagues deserve great thanks for introducing Ferrara into the 
conference calender. I can only hope it will have more appearances 
there in the future. 
This work has been supported by the NSF under the grant 
PHY 96-05080. 



\begin{thebibliography}{99}

\bibitem{ROOS}
B. Laurent, M. Roos, {\em Phys. Lett.} {\bf 13} (1964) 269; 
{\em ibid.} {\bf 15} 104. 

\bibitem{PAIS} 
For an exciting historical account, see A. Pais, 
`CP violation: the first 25 years', {\em CP Violation in 
Particle and Astrophysics}, J. Tran Than Van (ed.), 
Edition Frontieres, 1989. 

\bibitem{CPBOOK}
For a more detailed recounting, see: 
I.I. Bigi, A.I. Sanda, {\em CP Violation}, Cambridge University 
Press, 2000. 

\bibitem{KRAMERS}
For a more recent reference, see: A. Messiah, 
{\em Quantum Mechanics}, Vol. II, North-Holland, Amsterdam, 1965, 
p. 675. 

\bibitem{BS}  
I. Bigi, A.I. Sanda, {\em Nucl. Phys.} {\bf B 193} (1981) 85. 

\bibitem{MOHA}
R. Mohapatra, {\em Phys. Rev.} {\bf D 6} (1972) 2023. 

\bibitem{SUWE} 
L. Wolfenstein, {\em Phys. Rev. Lett.} {\bf 13} (1964) 562.

\bibitem{KM} 
M. Kobayashi, T. Maskawa, {\em Prog. Theor. Phys.} {\bf 49} 
(1973) 652. 

\bibitem{ITEP}
A. Vainshtein, V. Zakharov, M. Shifman, 
{\em Sov. Phys. JETP} {\bf 45} (1977) 670.

\bibitem{GILMAN}
F. Gilman, M. Wise, {\em Phys. Rev.} {\bf D 20} (1979) 2392.

\bibitem{BURAS} 
A. Buras, P. Gambino, M. Gorbahn, S. Jager, L. Silvestrini, 
hep-ph/0007313. 


\bibitem{NIU}
K. Niu, E. Mikumo, Y. Maeda, {\em Prog. Theor. Phys.} {\bf 46} 
(1971) 1644. 

\bibitem{PECCEI} 
For a clear review, see: R. Peccei, in: 
{\em CP Violation}, C. Jarlskog (ed.), World Scientific, Singapore, 
1988. 

\bibitem{CARTER} 
A. B. Carter, A.I. Sanda, {\em Phys. Rev.} {\bf D 23} (1981) 1567. 



\bibitem{DOK}
I. Bigi, Y. Dokhshitzer, V. Khoze, J. K\" uhn, P. Zerwas, 
{\em Phys. Lett.} {\bf B 181} (1986) 157. 

\bibitem{HQT} 
I.I. Bigi, M.Shifman, N. Uraltsev, {\em Annu. Rev. Nucl. 
Part. Sci.} {\bf 47} (1997) 591, 
with references to earlier work.  

\bibitem{SONI} 
A. Soni, these Proceed. 

\bibitem{KENWAY} 
R. Kenway, Invited Plenary Talk given at ICHEP2000, July 27 - 
August 2, 2000, Osaka, Japan, to appear in the Proc. 

\bibitem{LAT}
J. Simone et al., {\em Nucl. Phys. Proc. Supp.} {\bf 83} 
(2000) 334.


\bibitem{MEL}
K. Melnikov, A. Yelkhovsky, {\em Phys. Rev.} {\bf D 59} (1999) 114009. 

\bibitem{HOANG} 
A. Hoang, {\em Phys. Rev.} {\bf D 61} (2000) 034005. 

\bibitem{SIGNER}
M. Beneke and A. Signer, {\em Phys. Lett.} {\bf B 471} (1999) 233.

\bibitem{URI1} 
I. Bigi, hep-ph/9907270. 

\bibitem{BABARBOOK}
The BABAR Physics Book, P. Harrison \& H. Quinn (eds.), SLAC-R-504. 

\bibitem{OSAKA} 
I.I. Bigi, Invited Plenary Talk given at ICHEP2000, July 27 - 
August 2, 2000, Osaka, Japan, to appear in the Proc., 
hep-ph/0009021.

\bibitem{CINABRO} 
D. Cinabro, Invited Plenary Talk given at ICHEP2000, July 27 - 
August 2, 2000, Osaka, Japan, to appear in the Proc.

\bibitem{OPAL}
The OPAL Collab., preprint CERN-EP-2000-100. 

\bibitem{FABB} 
M. Fabbrichesi, these Proceed.


\bibitem{PARODI} 
F. Parodi, these Proceed. 

\bibitem{DELTARULE} 
M. Gell-Mann, A. Pais, {\em Proc. Glasgow Conf. Nuclear and 
Meson Physics}, E.H. Bellamy and R.G. Moorhouse (eds.), 
Pergamon Press, Oxford, 1955.

\bibitem{SANDA} 
T. Morozumi, C.S. Lim and A.I. Sanda, 
{\em Phys. Rev. Lett.} 
{\bf 65}(1990) 404. 



\bibitem{TRIESTE} 
S. Bertolini, J. Eeg, M. Fabbrichesi, {\em Rev. Mod. Phys.} 
{\bf 72} (2000) 65.  

\bibitem{TOD} 
A.I. Oksak, I.T. Todorov, 
{\em Commun. Math. Phys.} {\bf 9} (1968) 2146. 

\bibitem{TUMB} 
I.I. Bigi, {\em Z. Phys.} {\bf C 12} (1982) 235.  

\bibitem{KOST} 
A. Kostelecky, preprint hep-ph/9912528; 
M. Kobayashi, A.I. Sanda, {\em Phys. Rev. Lett.} 
{\bf 69}(1992) 3139. 

\bibitem{EPR} 
A. Einstein, B. Podolsky, N. Rosen, 
{\em Phys. Rev.} {\bf D 47}(1935) 777. 



\bibitem{CPLEAR} 
A. Apostolakis et al., {\em Phys. Lett.} {\bf B 444} (1998) 43; 
{\em Phys. Lett.} {\bf B 456} (1999) 297.  

\bibitem{SEHGAL} 
L. Sehgal, M. Wanninger, {\em Phys. Rev.} {\bf D 46}(1992) 1035; 
5209 (E). 


\bibitem{TBS} 
I.I. Bigi, A.I.  Sanda, {\em Phys. Lett.} {\bf B 466} (1999) 33.

\bibitem{TRIANGLES} 
I.I. Bigi, A.I. Sanda, hep-ph/9909479. 

 







 
































\bibitem{BUOSC} 
I. Bigi, N. Uraltsev, hep-ph/0005089, to appear in 
{\em Nucl. Phys.} {\bf B}.  

\bibitem{PETROV} 
S. Bergmann, Y. Grossman, Z. Ligeti, Y. Nir, A. Petrov, 
{\em Phys. Lett.} {\bf B 486} (2000) 418.




 


\bibitem{MNS} 
Z. Maki, M. Nakagawa, S. Sakata, {\em Prog. Theor. Phys.} {\bf 30} 
(1963) 727.







\end{thebibliography}
\end{document}